\documentclass[preprints,article,accept,pdftex,moreauthors,galaxies]{Definitions/mdpi} 
\usepackage{amsmath,mathtools,mathrsfs, bm}
\usepackage{verbatim}
\usepackage{xspace}
\usepackage{aas_macro}
\usepackage{multirow}

\usepackage[usestackEOL]{stackengine}
\edef\tmp{\the\baselineskip}
\setstackgap{L}{\tmp}

\def\lsim{\mathrel{\raise.3ex\hbox{$<$\kern-.75em\lower1ex\hbox{$\sim$}}}}
\def\gsim{\mathrel{\raise.3ex\hbox{$>$\kern-.75em\lower1ex\hbox{$\sim$}}}}

\def\muas{\mu{\rm as}} %....micro-arcseconds
\def\uas{$\mu$as\xspace} % uas
\def\m87{M87$^*$\xspace}
\def\sgra{Sgr~A$^*$\xspace}

\def\comrade{\texttt{Comrade.jl}\xspace}
\def\mring{m-ring\xspace}
\def\tmring{thick m-ring\xspace}
\def\stmring{stretched thick m-ring\xspace}
\def\Jy{\mathrm{Jy}}
\def\ehtim{\texttt{eht-imaging}\xspace}

\firstpage{1} 
\pubvolume{1}
\issuenum{1}
\articlenumber{0}
\pubyear{2022}
\copyrightyear{2022}
\datereceived{} 
\dateaccepted{} 
\datepublished{} 
\hreflink{https://doi.org/} % If needed use \linebreak
\Title{Measuring Photon Rings with the ngEHT}

\TitleCitation{ngEHT Photon Rings}

\Author{
Paul~Tiede$^{1,2,*}$\orcidA{}, 
Michael~D.~Johnson$^{1,2}$\orcidB{},
Dominic~W.~Pesce$^{1,2}$\orcidC{}, 
Daniel~C.~M.~Palumbo$^{1,2}$\orcidD{},
Dominic~O.~Chang$^{1,2}$, 
and Peter~Galison$^{2,3,4}$\orcidF{}
}

\AuthorNames{Paul Tiede, Michael D. Johnson, Dominic W. Pesce, Daniel C. M. Palumbo, Dominic O. Chang, Peter Galison}

\AuthorCitation{Tiede, P.; Johnson, M.D.; Pesce, D.W.; Palumbo, D.; Chang, D.O.; Galison, P.}
\address{%
$^{1}$ \quad Center for Astrophysics $|$ Harvard \& Smithsonian, 60 Garden Street, Cambridge, MA 02138, USA\\
$^{2}$ \quad Black Hole Initiative at Harvard University, 20 Garden Street, Cambridge, MA 02138, USA\\
$^{3}$ \quad Department of Physics, Harvard University, Cambridge, MA 02138, USA\\
$^{4}$ \quad Department of History of Science, Harvard University, Cambridge, MA 02138, USA
}

\corres{Correspondence: paul.tiede@cfa.harvard.edu}

\abstract{
General relativity predicts that images of optically thin accretion flows around black holes should generically have a ``photon ring,'' composed of a series of increasingly sharp subrings that correspond to increasingly strongly lensed emission near the black hole. Because the effects of lensing are determined by the spacetime curvature, the photon ring provides a pathway to precise measurements of the black hole properties and tests of the Kerr metric. We explore the prospects for detecting and measuring the photon ring using very long baseline interferometry (VLBI) with the Event Horizon Telescope (EHT) and the next generation EHT (ngEHT). We present a series of tests using idealized self-fits to simple geometrical models and show that the EHT observations in 2017 and 2022 lack the angular resolution and sensitivity to detect the photon ring, while the improved coverage and angular resolution of ngEHT at 230 GHz and 345 GHz is sufficient for these models. 
We then analyze detection prospects using more realistic images from general relativistic magnetohydrodynamic simulations by applying ``hybrid imaging,'' which simultaneously models two components: a flexible raster image (to capture the direct emission) and a ring component. Using the Bayesian VLBI modeling package \comrade, we show that the results of hybrid imaging must be interpreted with extreme caution for both photon ring detection and measurement --- hybrid imaging readily produces false positives for a photon ring, and its ring measurements do not directly correspond to the properties of the photon ring. 
}

\defcitealias{broderick_hybrid_2020}{B20}
\defcitealias{broderick_photon_2022}{B22}

\keyword{Black Holes; Photon Rings; Radio Astronomy; VLBI}

\begin{document}

\nocite{M87EHTCI}
\nocite{M87EHTCII}
\nocite{M87EHTCIII}
\nocite{M87EHTCIV}
\nocite{M87EHTCV}
\nocite{M87EHTCVI}
\nocite{M87EHTCVII}
\nocite{M87EHTCVIII}
\nocite{SgrAEHTCI}
\nocite{SgrAEHTCII}
\nocite{SgrAEHTCIII}
\nocite{SgrAEHTCIV}
\nocite{SgrAEHTCV}
\nocite{SgrAEHTCVI}

%%%%%%%%%%%%%%%%%%%%%%%%%%%%%%%%%%%%%%%%%%

\section{Introduction}
\label{sec:intro}

\begin{figure}[!t]
    \centering
    \includegraphics[width=\textwidth]{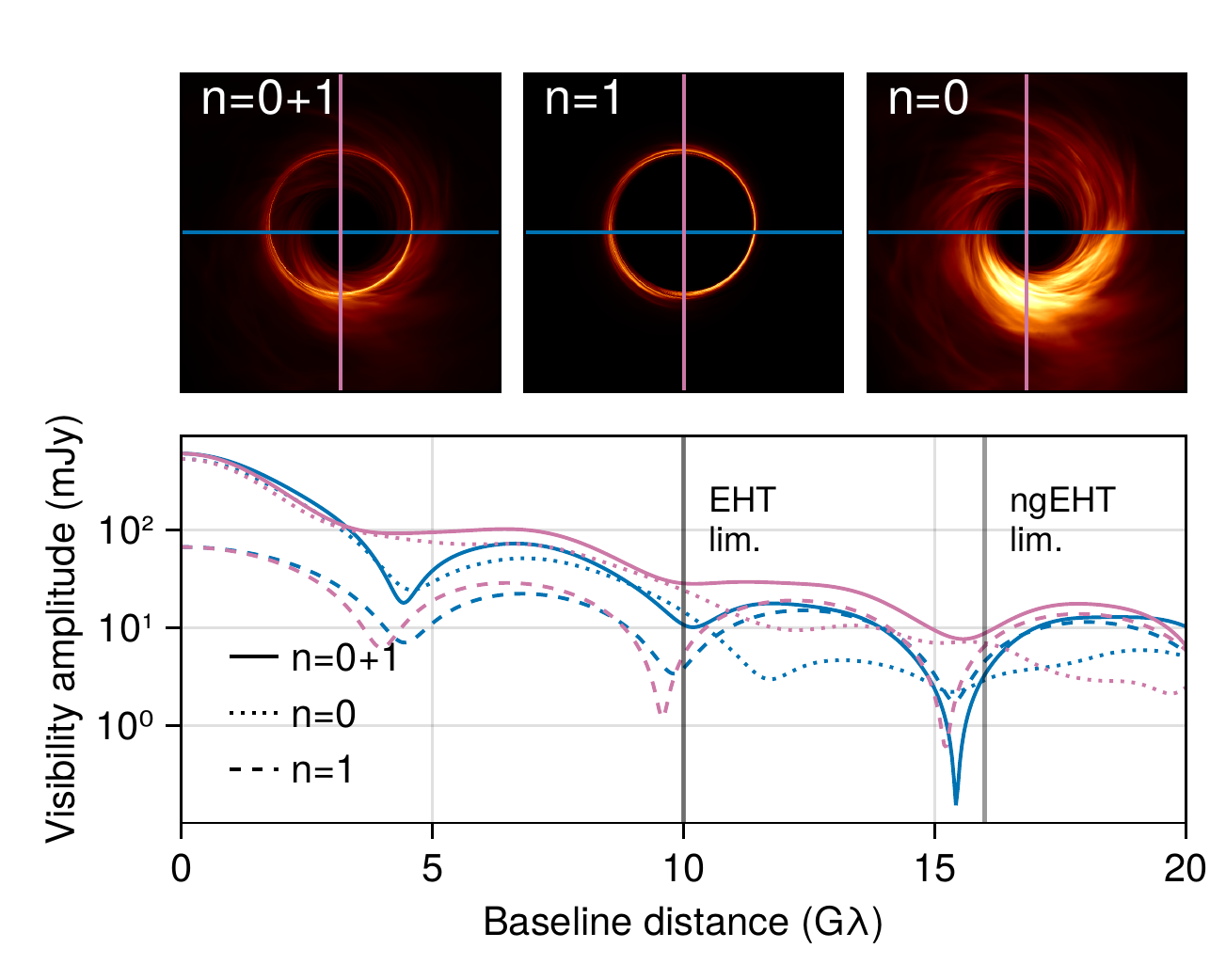}
    \caption{The image of a black hole can be decomposed into subimages that are indexed by the number of half orbits that their photons traveled around the black hole before reaching the observer. In this scheme, the $n=0$ emission (top right panel) is the ``direct'' image of the accretion flow and is dominated by astrophysical emission structure. The $n=1$ emission (top middle panel) is the ``secondary'' image, consisting of photons that have traveled a half orbit around the black hole before reaching the observer. The actual observed image is a sum of all $n$ subimages (top right panel). The bottom panel shows visibility amplitudes of these (sub)images for projected baselines that are parallel (blue) and perpendicular (pink) to the black hole spin axis. The longest EHT and ngEHT baselines, indicated with vertical black lines, occur at baseline lengths for which the $n=0$ and $n=1$ contributions are comparable, raising the prospect of distinguishing them through modeling.}
    \label{fig:photon_decomp}
\end{figure}

Simulated images of optically thin accretion flows around supermassive black holes (SMBHs) generically exhibit a nested series of ``photon rings'' produced from strong gravitational lensing of photon trajectories near the black hole \citep[e.g.,][]{M87EHTCV,SgrAEHTCV}.  These increasingly sharp ring-like features are exponentially demagnified as they converge on an asymptotic critical curve \citep{Bardeen_1973,Luminet_1979}, and they can be indexed by the number $n$ of half-orbits that light takes around the black hole, as shown in \autoref{fig:photon_decomp} \citep{Darwin_1959,Gralla_2019,Johnson_2020}.  
Because the null geodesics that define the photon ring are determined by the spacetime curvature and are negligibly affected by accreting plasma, detection of an $n > 0$ photon ring would provide striking evidence that the supermassive compact objects in galactic cores are Kerr black holes and would provide a pathway to precisely measuring their properties.

To date, measurements of the horizon-scale emission structure around black holes %remain the domain of 
are only possible using 
millimeter-wavelength very long baseline interferometry (VLBI).  The Event Horizon Telescope (EHT) is a globe-spanning network of (sub)millimeter radio telescopes that has carried out VLBI observations of the SMBHs \m87 and \sgra on horizon scales \citep{M87EHTCI,M87EHTCII,M87EHTCIII,M87EHTCIV,M87EHTCV,M87EHTCVI,M87EHTCVII,M87EHTCVIII,SgrAEHTCI,SgrAEHTCII,SgrAEHTCIII,SgrAEHTCIV,SgrAEHTCV,SgrAEHTCVI}.  The next-generation EHT (ngEHT) plans to build on the capabilities of the EHT by adding multiple new telescopes to the array, increasing the frequency coverage, and improving the sensitivity by observing with wider bandwidths \citep{Doeleman_2019}.  Though the ngEHT will operate with an unprecedentedly fine diffraction-limited angular resolution of $\sim$15\,\uas, the $n=1$ photon ring is anticipated to be finer still; the expected thickness of the $n=1$ photon ring in \m87 corresponds to an angular size of less than ${\sim}4$\,\uas \citep{Johnson_2020}.  Direct imaging of the $n=1$ photon ring will thus likely remain unachievable for the foreseeable future, and studies of this feature using ground-based VLBI will require some degree of ``superresolution'' via judicious application of parameterized models of the source structure.

At least two classes of modeling methodology currently show promise for extracting superresolved photon ring signatures from VLBI measurements of black holes: models that parameterize the three-dimensional distribution of the material in the vicinity of the black hole \citep[e.g.,][]{Broderick_2009,Tiede_2020,Palumbo_2022_KerrBAM}, and models that parameterize the two-dimensional distribution of the emission morphology as seen on the sky \citep{broderick_hybrid_2020,broderick_photon_2022}. In either case, because the additional information supplied by the model specification is supporting the extraction of superresolved structural information, it is important to quantify precisely what defines a photon ring ``detection.''  For instance, the most compelling detection might not require the assumption that general relativity (GR) is true, while a somewhat weaker claim of detection might test for the presence of this feature under the assumptions of GR. Likewise, methods could utilize models that \textit{assume} the existence of the photon ring to make measurements of black hole parameters without needing to meet potentially more stringent criteria for an unambiguous \textit{detection} of the same feature.

A parameterized modeling approach to study the photon ring was recently developed by \citet{broderick_hybrid_2020} (hereafter \citetalias{broderick_hybrid_2020}), who employ a ``hybrid imaging'' technique that fits a thin geometric ring component alongside a more flexible pixel-based image component, where the pixel fluxes are treated as model parameters. \citet{broderick_photon_2022} (hereafter \citetalias{broderick_photon_2022}) applied this technique to the EHT observations of \m87, finding that the diameter of the thin ring component is well-constrained by the EHT data; the authors associate this component with the $n=1$ photon ring.  While the value and stability of the diameter of this component across different datasets support its identification as %the $n=1$ ring
an image feature that is determined by the spacetime, other aspects -- particularly the fraction of the total flux density that is recovered in the thin ring -- challenge its association with the $n=1$ ring. This ambiguity underscores the need to quantify exactly what constitutes a photon ring detection.

In this paper, we aim to investigate the efficacy of tools such as hybrid imaging to extract photon ring signatures from EHT- and ngEHT-like data and to determine what VLBI measurements are necessary and sufficient to reliably detect a photon ring.  In \autoref{sec:geom_modeling}, we conduct tests using simple geometric models, deriving necessary conditions to detect the $n=1$ photon ring.  Next, in \autoref{sec::hybrid_modeling}, we explore the application and limitations of the hybrid imaging approach to detect and measure the photon ring, and we perform tests using more realistic synthetic data from general relativistic magnetohydrodynamic (GRMHD) simulations.  In \autoref{sec:discussion}, we summarize these results and discuss their implications for the EHT, ngEHT, and other future VLBI arrays.

\section{Geometric Modeling}\label{sec:geom_modeling}

We begin with a series of idealized tests, generating simulated data from a simple geometric on-sky model that includes a proxy for the photon ring, and then fitting the same model to these data. This so-called \textit{self-fit} procedure guarantees that model parameter posteriors are directly interpretable. However, the clarity of this procedure comes with the penalty of being artificially optimistic; 
it provides requirements for detecting the photon ring that are likely necessarily but almost certainly not sufficient. 
Hence, if these self-fits to simulated data cannot detect a photon ring with a given array, then we expect that photon ring detection with the same array in realistic settings will be impossible. 

The structure of this section is as follows. First, we describe our geometric model (\autoref{sec::Geometric_Model}). Next, we outline our construction of simulated data and the fitting procedure (\autoref{ssec:geom_setup}). Finally, we perform self-fits for a variety of EHT and ngEHT arrays to assess the requirements for detecting the $n=1$ photon ring (\autoref{sec:geom_results}).

\subsection{Specifying the Geometric Model}
\label{sec::Geometric_Model}

Our simple parametric model is motivated by the expected image structure for optically thin emission near a black hole consisting of multiple ring-like structures. For each component, we use the \mring model from \citet{Johnson_2020} and \citet{SgrAEHTCIV}. This model is an infinitesimally thin ring with azimuthal brightness modulation determined by angular Fourier coefficients, which is then convolved with a Gaussian kernel $G$. We restrict ourselves to a first-order Fourier expansion, giving the following intensity profile for the thin ring:
\begin{equation}\label{eq:mring}
    M(r, \theta| d_i, a_i, b_i, F_i) = \frac{F_i}{\pi d_i}\delta(r - d_i/2)\left(1 + a_i\cos(\theta) - b_i\sin(\theta)\right),
\end{equation}
where we parameterize $a_i,b_i$ using a polar representation $a_i = A_i \cos \phi_i$ and $b_i = A_i \sin \phi_i$, where $A_i$ is the amplitude and $\phi_i$ is the phase of the first-order Fourier coefficient. Finally, $F_i$ and $d_i$ are the flux and diameter of the ring, respectively. Note that we have included a subscript, $i$, in anticipation of the nested photon rings. The location of the observed $n$ photon rings relative to the emitting plasma are shifted as a function of spin and inclination. Therefore, we allow the centroid of the rings to be displaced by an amount $x_i,\, y_i$. To give the ring finite width, we convolve the \mring with a symmetric Gaussian:
\begin{equation}
    G(r, \theta| w_i) = \frac{4\log(2)}{\pi w_i^2}\exp\left(-\frac{4\log(2)r^2}{w_i^2}\right)
\end{equation}
where $w_i$ is the Gaussian's full width at half maximum (FWHM).  We denote the thick \mring model by $T(x,y) = M\star G$, where $\star$ is the convolution operator. Finally, the shape of the ring is also of interest since it encodes information about the spin and inclination of the central black hole \citep[see, e.g.,][]{Takahashi_2004,Johannsen_2010, Broderick_2014,Medeiros_2020,Johnson_2020,Farah,GrallaPhotonRingShape}. To add ring ellipticity, we modify the intensity map of the thick \mring \begin{equation}
    T(x,y) \to R_{\xi}T(x, (1+\tau)y) = I(x,y)
\end{equation}
where $R_{\xi}$ rotates the image by $\xi$ radians counter-clockwise, and $\tau > 0$ parametrizes the ring ellipticity. Formally, $\tau$ is related to the eccentricity $e$ of the elliptical (stretched) ring via $e = \sqrt{1-1/(1+\tau)^2} \approx \sqrt{2\tau}$. We denote this model by $I$ and call it the \stmring.  

The \stmring forms the base image for each nested photon ring. The final model that we use is a sum of multiple \stmring components:
\begin{equation}\label{eq:geom_model}
    I^{0:N}(x,y) = \sum_{n=0}^N I(x,y | F_i, d_i, w_i, A_i, \phi_i, \tau_i, \xi_i, x_i, y_i).
\end{equation}

\begin{table}[!t]
    \centering
    \caption{Arrays used for synthetic data. For additional details on EHT sites, see \citep{M87EHTCII}; for additional details on ngEHT sites, see \citep{Raymond2021}. Note that the SPT cannot observe \m87 so does not contribute to the tests shown in this paper. New ngEHT phase~1 sites use specifications for existing facilities (HAY: 37-m, OVRO: 10.4-m) and are 6.1-m for new locations (BAJA, CNI, LAS); new ngEHT phase~2 sites assume 8-m diameters with the exception of the AMT, which is planned to be 15-m \citep{Backes_2016}.}
    \begin{tabular}{c|c|l}
        Array & Freq.~(GHz) & Sites \\
    \hline
    \hline
        EHT 2017 & $230$ & (8) ALMA, APEX, JCMT, LMT, IRAM, SMA, SMT, SPT \\
        \hline
        EHT 2022 & $230$ & (11) EHT 2017, KP, NOEMA, GLT\\ 
        \hline
        ngEHT phase 1 & $230,\,345$ & (16) EHT 2022, BAJA, CNI, HAY, LAS, OVRO \\
        \hline
        ngEHT phase 2 & $230,\,345$ & (22) ngEHT phase 1, GARS, AMT, CAT, BOL, BRZ, PIKE \\
    \end{tabular}
    \label{tab:arrays}
\end{table}

\subsection{Simulated Observations and Fitting Procedure}\label{ssec:geom_setup}

To create simulated data, we use \autoref{eq:geom_model} with \mring parameters motivated by the observed structure and expected gravitational lensing of \m87 \citep{M87EHTCVI}. Because we are focused on distinguishing the $n=0$ and $n=1$ structure, our model for the construction of the simulated data consists of two nested rings (i.e., $N=1$) with equal brightness. Their diameters $d_i$ are related by 
\begin{equation}\label{eq:diam}
    d_1 = 2\rho_{\rm c} + (d_0 - 2\rho_{\rm c})e^{-\gamma}.
\end{equation}
We set $\rho_{\rm c} = 19\,\muas$ and $\gamma = \pi$, which approximates the structure of the photon ring in \m87 given its mass \citep{M87EHTCVI} and low viewing inclination \citep{Walker_2018}. Additionally, the width of the rings will be given by 
\begin{equation}\label{eq:width}
    w_1 = w_0 e^{-\gamma},
\end{equation} 
and the flux by 
\begin{equation}\label{eq:flux}
    F_{n+1}/F_n \sim e^{-\gamma}.
\end{equation}
Note that these expressions are a rather crude approximation of the precise structure expected in black hole images. Nevertheless, given that we want to explore the potential for photon ring detections that do not explicitly assume general relativity and that our goal is only to define a minimum threshold for detection, we do not regard this as a significant limitation. 

Our complete model description is as follows:
\begin{itemize}
    \item {\bf Flux:} $F_0 = 0.6\,{\rm Jy}$, and $F_1=e^{-\pi}F_0=0.03\,\Jy$ (\autoref{eq:flux})
    \item {\bf Diameter:} $d_0 = 45\,\muas$, and $d_1 = 2\rho_{\rm c} + (d_0 - 2\rho_{\rm c})e^{-\pi} = 38.3\,\muas$ (\autoref{eq:diam})
    \item {\bf Width:} $w_0 = 18\,\muas$, and $w_1 = w_0 e^{-\pi} = 0.78\,\muas$ (\autoref{eq:width})
    \item {\bf Brightness Asymmetry:} $A_0 = A_1 = 0.15$, and $\phi_0 = \phi_1 = -\pi/2$
    \item {\bf Ellipticity:} $\tau_0 = \tau_1 = 0.05$, and $\xi_0 = \xi_1 = \pi/3$
    \item {\bf Ring Centers:} $x_0 = y_0 = x_1 = y_1 = 0\,\muas$
\end{itemize}
The image structure of this model is shown in the leftmost panel of \autoref{fig:geom_recon}. 

We generate simulated data using \ehtim \citep{Chael_2018} and the \texttt{ngehtsim}\footnote{\url{https://github.com/Smithsonian/ngehtsim}} package. This procedure integrates historical weather data at each site using similar methods to \citet{Raymond2021}. For the EHT arrays, we use 2\,GHz of recorded bandwidth with dual polarization, while for the ngEHT arrays, we use 8\,GHz of recorded bandwidth with dual polarization.  To approximate the practical limitations related to fringe detection, we remove baselines that are not connected to another baseline with an ${\rm SNR} > 5$ within a 10-second integration time. Finally, we segment and coherently average the data over 5-minute intervals, emulating standard on-sky scans. For each 5-minute scan, we add station-based gain corruptions, which add 10\% Gaussian multiplicative Gaussian noise in amplitudes and uniform $[0, 2\pi)$ noise in visibility phases. We include thermal noise but do not include any additional non-closing errors, such as polarimetric leakage.

To extract model parameters from the simulated data, we use Bayesian inference. Our goal is to find the posterior distribution $p({\bm\theta} | {\bm D})$ for our model parameters $\bm \theta$ given the data products $\bm D$:
\begin{equation}
    p({\bm \theta} | {\bm D}) = \frac{p({\bm D} | {\bm \theta}) p({\bm \theta})}{p({\bm D})}.
\end{equation}
The $p({\bm D}|{\bm \theta})$ distribution is often called the likelihood and is sometimes denoted by $\mathcal{L}({\bm D}|{\bm \theta})$, $p({\bm \theta})$ is the prior, and $p({\bm D})$ is the marginal likelihood or evidence. 

In this work, we use log-closure amplitudes and closure phases as our data products.\footnote{We have also explored fitting other data products, including visibility amplitudes and complex visibilities, and we find that our conclusions are unchanged.} The benefit of closure products is that they are immune to station-based gain errors (such as those introduced in the synthetic data generation). However, the likelihood functions of closure quantities are non-Gaussian in the low-SNR limit \citep{TMS,Blackburn_2020}. For this paper, we used the high-SNR expression from \citet{SgrAEHTCIV} and flagged any closure products that had ${\rm SNR} < 3$. For the EHT and ngEHT 230~GHz observations, this cut removed 1\% of the closure products. For the ngEHT 345~GHz observations, this cut removed 25\% of the closure products.

For model fitting, we use the same model prescription as above, but we parameterize the width and total flux density as follows:
\begin{itemize}
    \item $w_1 = \epsilon_1 w_0$ (the width of the $n=1$ ring is forced to be thinner than the first)
    \item $F_0 + F_1 = 1\,\rm Jy$ (the total flux density is forced to be unity)
\end{itemize}
Both of these choices are for computational efficiency when sampling the posterior. The first prevents a trivial label-swapping degeneracy between the two m-rings, and the second resolves the trivial flux-rescaling degeneracy that occurs when using only closure products. Additionally, we force the $x_0 = y_0 = 0\,\muas$ to provide a phase center for the reconstructions. For multi-frequency observations with the ngEHT, we assume a flat spectral index (i.e., we use the same model to fit both 230 and 345\,GHz observations and also for the ground-truth model). \autoref{tab:priors} lists the priors assumed for each model parameter.

\begin{table}[!h]
    \centering
    \begin{tabular}{c|c}
        Parameter & Prior \\
        \hline \hline
        $d_{0,1}$ & $\mathcal{U}(20~\muas, 60~\muas)$ \\
        $w_0$ & $\mathcal{U}(0.1~\muas, 40~\muas)$ \\
        $\epsilon_1$ & $\mathcal{U}(0.0, 1.0)$\\
        $A_{0,1}$        & $\mathcal{U}(0.0, 0.5)$\\
        $\phi_{0,1}$   & $\mathcal{U}(-\pi, \pi)$\\
        $\tau_{0,1}$     & $\mathcal{U}(0.0, 0.5)$ \\
        $\xi_{0,1}$     & $\mathcal{U}(0, \pi)$ \\
        $y_1$   & $\mathcal{U}(-10~\muas, 10~\muas)$ \\
        $x_1$   & $\mathcal{U}(-10~\muas, 10~\muas)$ \\
    \end{tabular}
    \caption{Priors used for the geometric self-fits with two \mring components. $\mathcal{U}(a,b)$ represents the uniform distribution on the interval $(a,b)$.}
    \label{tab:priors}
\end{table}

To fit the model to the simulated data, we use the Bayesian VLBI modeling package \comrade \citep{comrade} implemented within the Julia programming language \cite{Julia-2017}.  To sample from the posterior, we first use the pathfinder algorithm \citep{pathfinder} and its Julia implementation \texttt{Pathfinder.jl} \citep{seth_axen_pathfinder} to find a Gaussian approximation of the posterior. This approximation tends to be poor, but it helps initialize MCMC sampling methods that enable more precise estimates of the posterior. To sample from the posterior, we use the NUTS algorithm \citep{hoffman2014no} and its Julia implementation \citep{xu2020advancedhmc}. To initialize our sampling, we draw a random sample from the pathfinder variational approximation, and we use the diagonal elements of the covariance matrix to initialize the NUTS mass matrix. We find that this greatly reduces the amount of time required for NUTS to adapt to our posterior. We run the NUTS sampler for 3000~adaptation steps and 5000~sampling steps. To check MCMC convergence, we compute the effective sample size of each chain (after removing the adaptation steps) and found $> 500$ effective samples for all model parameters. Note that HMC struggles to explore multi-modal posteriors, making it possible that we are missing parts of the posterior distribution. To test for multi-modality, we ran an optimizer from many starting locations. Other than pathological cases where the fit quality is extremely poor, we find no evidence for a multi-modal posterior.

\subsection{Results}
\label{sec:geom_results}

\begin{figure}
    \centering
    \includegraphics[width=\linewidth]{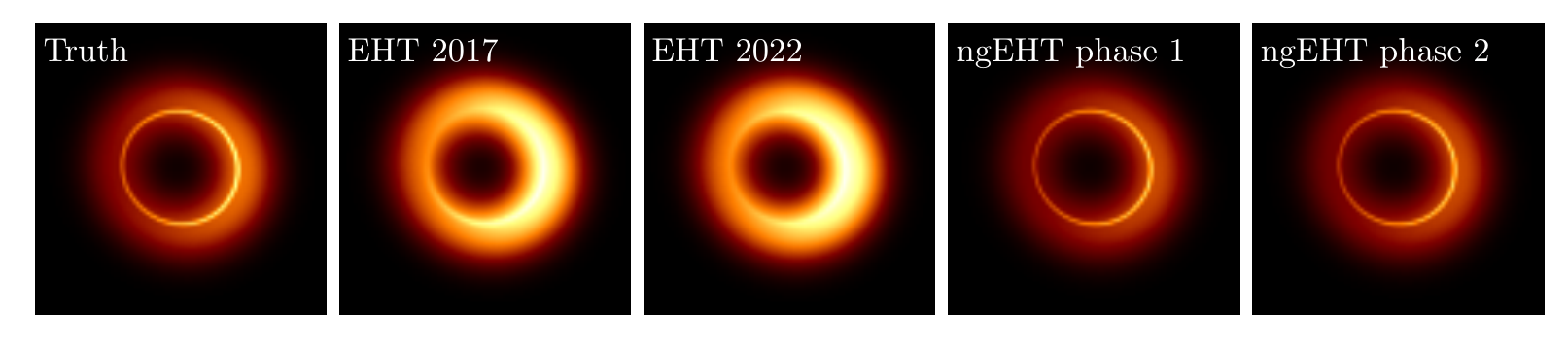}
    \caption{Results from self-fits of geometric models to synthetic data for a series of EHT and ngEHT arrays. Each fit includes two \mring components, as described in \autoref{sec::Geometric_Model}. The left panel shows the ground truth model; the remaining panels show the mean image for the array noted at the top. Even for this optimistic test, the EHT coverage is insufficient to identify the $n=1$ photon ring. However, the longer baselines of the ngEHT allow clear detection of the $n=1$ photon ring.}
    \label{fig:geom_recon}
\end{figure}

\begin{figure}
    \centering
    \includegraphics[width=\linewidth]{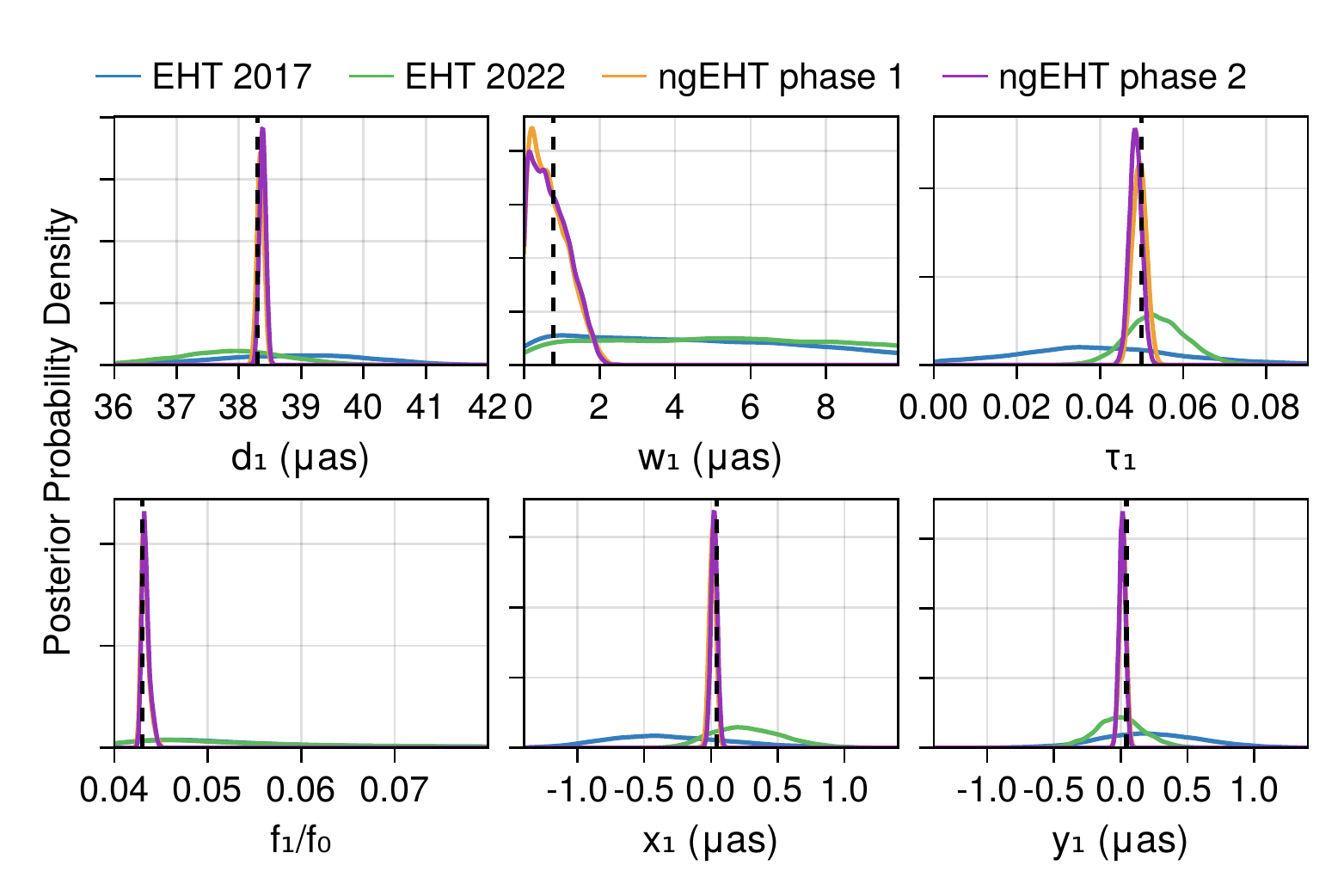}
    \caption{Geometric self-fit marginal posterior distributions for the examples shown in \autoref{fig:geom_recon}. The curves show the marginal posterior distributions for $n=1$ ring parameters of interest: diameter, width, ellipticity, fractional flux, and relative displacement. The black dashed line shows the true value for the $n=1$ ring component. Overall, we find that the EHT~2017 and 2022 coverage is insufficient to fully constrain the second ring. In particular, the width of the $n=1$ ring component is the most difficult quantity to measure.}
    \label{fig:width_posterior}
\end{figure}

\autoref{fig:geom_recon} shows the mean image estimated using synthetic data for each array in \autoref{tab:arrays}, while \autoref{fig:width_posterior} shows the marginal posteriors for the parameters of the secondary ring component. These results show that the secondary ring parameters have a hierarchy of measurement difficulty. For instance, all the arrays provide tight constraints on the ring diameter, flux, and centroid. The ring asymmetry is more challenging to constrain; the EHT 2017 coverage is inadequate to constrain the $n=1$ ring asymmetry\footnote{\citet{Tiede_2022b} find that the EHT 2017 coverage cannot even constrain the asymmetry of the $n=0$ image, $\tau_0$.}, while the EHT 2022 coverage is sufficient to constrain asymmetry to the ${\sim}2\%$ level, and the ngEHT coverage provides further improvement. The most challenging parameter to constrain is the width of the secondary ring, which is weakly constrained by both EHT arrays but is tightly constrained for the ngEHT arrays. Note that we used the combined $230$ and $345\, \rm GHz$ coverage for the ngEHT arrays; if restricted to just $230$ or $345$\,GHz, the arrays no longer constrain the width of the \mring to be less than $M/D \approx 4\,\muas$. Thus, these tests demonstrate the importance of ngEHT observations at both frequencies for \m87. 

Interestingly, the width posterior for ngEHT coverage does not appreciably change when moving from phase~1 to phase~2. This result arises because the six additional dishes in phase~2 bring more complete baseline coverage but do not increase the maximum baseline length or the SNR on long baselines, both of which are crucial for measuring the properties of the photon ring. Similarly to the marginal posterior for the width, the ellipticity posterior also shrinks considerably when moving from the EHT to the ngEHT arrays. Both ngEHT arrays measure the ellipticity to ${\lsim}\,1\%$ precision.

These idealized tests show that the EHT cannot detect the photon ring. However, the ngEHT may be able to meaningfully constrain its size, width, flux ratio, and ellipticity. 
Because the principal difficulty in a photon ring detection will be in distinguishing it from the direct emission, measuring each of these ring properties is important for claiming a detection of the photon ring. For instance, treating the ring thickness as a model parameter permits a crucial diagnostic of whether the data show a preference for a narrow ring, which is a key identifying property of the photon ring. Anomalous values in any of the fitted ring parameters may indicate that the direct emission affects the measured photon ring structure.

In the next section, we analyze the prospects for detecting the $n=1$ photon ring under more realistic circumstances, using images from GRMHD simulations to generate synthetic data and fitting models to these data using the hybrid imaging approach from \citetalias{broderick_hybrid_2020}.

\section{Hybrid Imaging}
\label{sec::hybrid_modeling}

We now explore more realistic tests of whether the EHT and ngEHT can detect and measure the photon ring, using synthetic data from GRMHD simulations and applying a flexible hybrid imaging approach from \citetalias{broderick_hybrid_2020} that we have implemented in \comrade. First, we review the original \citetalias{broderick_hybrid_2020} model and describe the modifications in our implementation of it (\autoref{ssec:hybrid_review}). Next, we apply hybrid imaging to a series of simulated datasets using the 2017~EHT array (\autoref{ssec:EHT2017}). Finally, we apply the hybrid model to simulated ngEHT data for the first time and assess the viability of photon ring measurements with the ngEHT and the hybrid imaging approach (\autoref{sec:hybrid_model_ngeht}).

\subsection{Review of Hybrid Imaging}\label{ssec:hybrid_review}
 
\citetalias{broderick_hybrid_2020} proposed modeling a black hole image using a decomposition consisting of two components. The first component is a rasterized image model given by
\begin{equation}\label{eq:raster}
    I(x, y) = \sum_{ij} I_{ij}\kappa(x - x_i)\kappa(y - y_j),
\end{equation}
where $\kappa(x)$\footnote{For this work we assume that $\kappa(x)$ has units of ${\rm sr}^{-1}$.} is the \textit{pulse} function that converts the raster of pixel fluxes $I_{ij}$ into a continuous image. For this work, we use a third-order B-spline kernel\footnote{\citetalias{broderick_hybrid_2020} uses a slightly different pulse function that doesn't preserve image positivity.}. The B-spline kernel is given by successive convolutions of the square wave pulse (${\rm Sq}(x)$) with itself; e.g., the third order B-spline is given by
\begin{equation}
    \kappa(x) = (\mathrm{Sq}\star \mathrm{Sq} \star \mathrm{Sq})(x),
\end{equation}
where ${\rm Sq}(x) = 1$ when $|x| < 1/2$ and $0$ otherwise.

The second model component in \citetalias{broderick_hybrid_2020} is a ring that is forced to be thin $(< 2~\muas)$, creating a natural scale separation in the model components. The authors suggested that this scale separation would allow the rasterized image model to predominantly fit the $n=0$ emission, while the ring component would predominantly fit the $n=1$ emission. Hence, the ring component would measure the properties of the photon ring. 

To test whether the hybrid imaging hypothesis works for the EHT data, \citetalias{broderick_hybrid_2020} analyzed mock data from five GRMHD simulations, with coverage and sensitivity corresponding to the 2017 EHT observations of \m87. They found that in 4/5 cases, the correct $n=1$ photon ring diameter was contained within the $95\%$ highest posterior density interval (HPDI) for the fitted ring diameter, and 5/5 models had the $n=1$ photon ring diameter within the $99\%$ HPDI. These results suggested that the measured diameter is correlated with the true $n=1$ photon ring diameter. However, \citetalias{broderick_hybrid_2020} found that the recovered ring flux was a factor of 2--3 times higher than the true $n=1$ photon ring flux; they argue that this excess flux matches expectations for an array with an angular resolution of $20\, \muas$. Namely, the flux approximately matches the integrated flux density within an annulus with a diameter of the $n=1$ photon ring and a width of $20\, \muas$. 

The \citetalias{broderick_hybrid_2020} analysis has two notable limitations: 1) it does not answer the question of whether the hybrid imaging approach will always favor placing a thin ring feature in the image regardless of the true on-sky appearance, and 2) it does not demonstrate that hybrid imaging can distinguish the $n=0$ and $n=1$ emission because the diameters of these components were very similar in all five tests. 
In the next section, we assess hybrid imaging for 2017 EHT data, focusing on tests that address both of these limitations.

\subsection{Testing Hybrid Imaging on EHT 2017 Data}\label{ssec:EHT2017}

\begin{figure}[t]
    \centering
    \includegraphics[width=\linewidth]{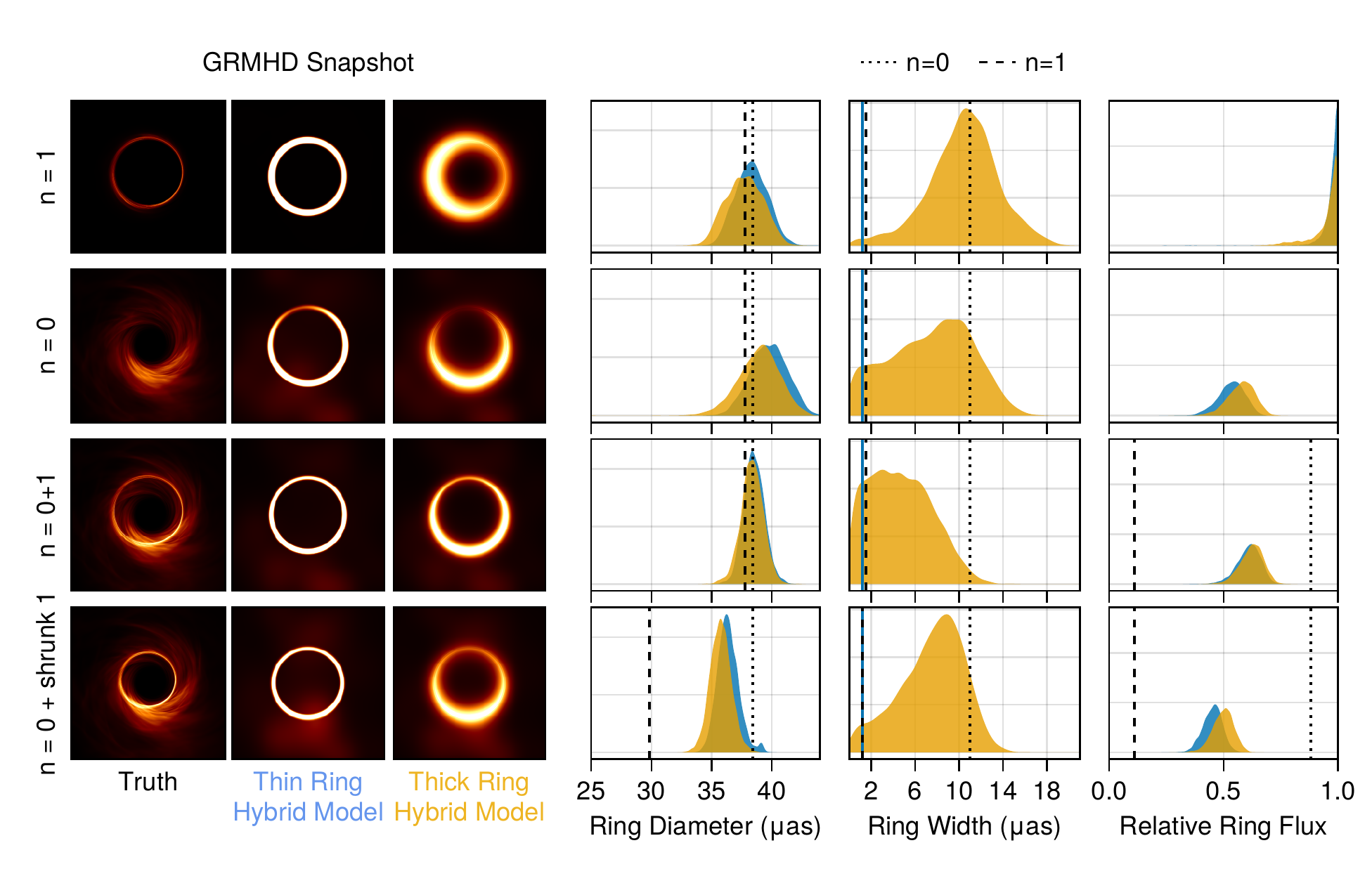}
    \caption{Results of fitting a GRMHD simulation snapshot with coverage corresponding to the EHT~2017 array. The GRMHD simulation has a spin of $0.9375$, inclination of $163^\circ$, $r_{\rm high} = 10$, and is in the MAD accretion state. The top row shows the results from fitting only the $n=1$ emission with the hybrid model, the second row shows fits to only the $n=0$ emission, and the third row shows fits to the combined $n=0+1$ emission. The bottom row shows fits to the combined $n=0+1$ emission after reducing the diameter of the $n=1$ ring by 21\%. All rows show both fits that force the ring component to be thin (blue) and fits that do not constrain the ring width (orange). The black dashed lines show true values for the $n=1$ photon ring, while the black dotted lines show the true values for the $n=0$ photon ring.}
    \label{fig:hybrid_fits}
\end{figure}

We first explore the application of hybrid imaging to synthetic data matching the EHT~2017 observations of \m87. For these tests, we generate data using the GRMHD simulation snapshot of \m87 shown in \autoref{fig:photon_decomp}. The GRMHD simulation was taken from \citet{Johnson_2020}, has a spin of $0.9375$, inclination of $163^\circ$, $r_{\rm high} = 10$, and is in the magnetically arrested accretion state (MAD). The GRMHD fluid simulations used in this paper were run using \texttt{iharm3d} \citep{Gammie_HARM_2003, IHARM3d_prather} and ray-traced using \texttt{ipole} \citep{IPOLE_2018}. The ray tracing decomposition into sub-images is taken from \citet{Palumbo_2022_pol}; for additional details on the snapshot generation pipeline, see \citet{Wong_2022}. To assess the ability of hybrid imaging to detect and/or measure the photon ring, we analyze four separate images: the first and second images contain just the $n=1$ and $n=0$ emission, respectively, and the third image is the combined $n=0+1$ emission, and the fourth image is similar to the $n=0+1$ image, but we have artificially shrunk the $n=1$ photon ring by $21\%$, corresponding to a $30~\muas$ diameter for the $n=1$ photon ring. The resulting images are shown in the first column of \autoref{fig:hybrid_fits}.

For each image, we generated simulated data whose properties matched the April~11 2017 EHT observations. We then fit two versions of the hybrid model described in the previous section for each dataset. Both models use an $8\times 8$ raster (see \autoref{eq:raster}) with a $90~\muas$ field of view and also a \tmring model. Note that we do not include ellipticity in the ring model fit to EHT 2017 data since we found that it was poorly constrained even for the geometric self-fits using EHT data (\citetalias{broderick_hybrid_2020} also did not include ellipticity).  

The difference between the two models is the prior on the thickness of the blurred \mring. The first model forces the FWHM of the ring to be $1\,\muas$ (similar to \citetalias{broderick_hybrid_2020}, who force the ring to be thin); we call this the \textit{thin-ring hybrid model}. For the second model, we also fit the thickness of the \tmring; we call this the \textit{thick-ring hybrid model}. The priors for the raster model are given by a Dirichlet prior with concentration parameter $\alpha=1$, effectively placing a uniform prior on the $N$ simplex where $N$ is the number of pixels in the image.\footnote{Our raster prior differs from the \citetalias{broderick_hybrid_2020} model which uses a log-uniform prior on the pixel intensity. \citetalias{broderick_photon_2022} also fits for the raster field of view and orientation.} For the ring component, we use the same priors as those for the first ring component in \autoref{tab:priors}, except for the thin-ring hybrid model where $w_0 = 1\,\muas$. We force the ring component and raster to be centered on the origin, which differs from \citetalias{broderick_hybrid_2020}, which fits for the ring centroid.

In our view, the two hybrid models serve different purposes, distinguished by their ability to fit all relevant ring parameters.
The thick-ring hybrid model makes fewer assumptions about the fitted ring component, making it a useful basis for \emph{detecting} a photon ring. 
%By fitting all of the ring parameters, we are not explicitly forcing a thin ring to be in the image. 
The thin-ring hybrid model imposes more assumptions about the fitted ring component, making it a useful basis for \emph{measuring} the remaining photon ring properties.  
\autoref{fig:width_posterior} provides motivation for forcing the ring to be thin, suggesting that the ring thickness is the most difficult parameter to constrain. 

To measure the ring diameters and widths for the $n=0$ and $n=1$ photon rings from the GRMHD images, we used the \texttt{VIDA.jl} package \citep{VIDA}. \texttt{VIDA} extracts image parameters by optimizing approximate template images that are parameterized by the features of interest. We used \texttt{VIDA}'s \texttt{SlashedGaussianRing} template, which provides estimates for the ring diameter, width, and brightness position angle. For our objective function, we used the Bhattacharyya divergence:
\begin{equation}
    \mathrm{Bh}(f|I) = \int \sqrt{f(x,y) I(x,y)}dx dy,
\end{equation}
where $f$ and $I$ are the template and image, respectively. To extract the ring parameters, we used \texttt{VIDA} on $n=0$ and $n=1$ images separately.

The results of the test are shown in \autoref{fig:hybrid_fits}. We find that the thick and thin ring models give similar results for the measured ring diameter and relative ring flux. This suggests that the thin ring component is not focusing on a different aspect of the image than the thick ring component. Comparing the $n=0$ and $n=0+1$ fits, we find that the results are very similar for both ring models.

Focusing on the thin ring model, we find that the measured diameter is $39.8^{+1.7}_{-0.7}~\muas$ for the $n=0$ fits and $38.5^{+1.0}_{-0.3}~\muas$ for the $n=1$ fits; the relative flux of the ring is $0.52^{+0.06}_{-0.03}$ for the $n=0$ fits and $0.61^{+0.05}_{-0.02}$ for the $n=0+1$ fits. Therefore, we find that the assumption of a thin ring does not appreciably change the estimated ring parameters. These results are consistent with the conclusions of \autoref{sec:geom_modeling}, which showed that the EHT~2017 array could not meaningfully constrain the thickness of a thin ring feature in \m87.

For the thin ring fits, we find that the model always places $\sim 50-60\%$ of the flux in the thin ring component. Additionally, the thin ring diameter of the $n=0+1$ image appears to be an average of the $n=0$ and $n=1$ fits, suggesting that its measured diameter is a combination of both. For the thick ring fits, we find that the measured flux and diameter of the m-ring component are very similar to the thin ring fits. The width for the thick ring analysis is very uncertain -- going from $0\, \muas$ to $15-20\, \muas$.

\begin{figure}[t]
    \centering
    \includegraphics[width=\linewidth]{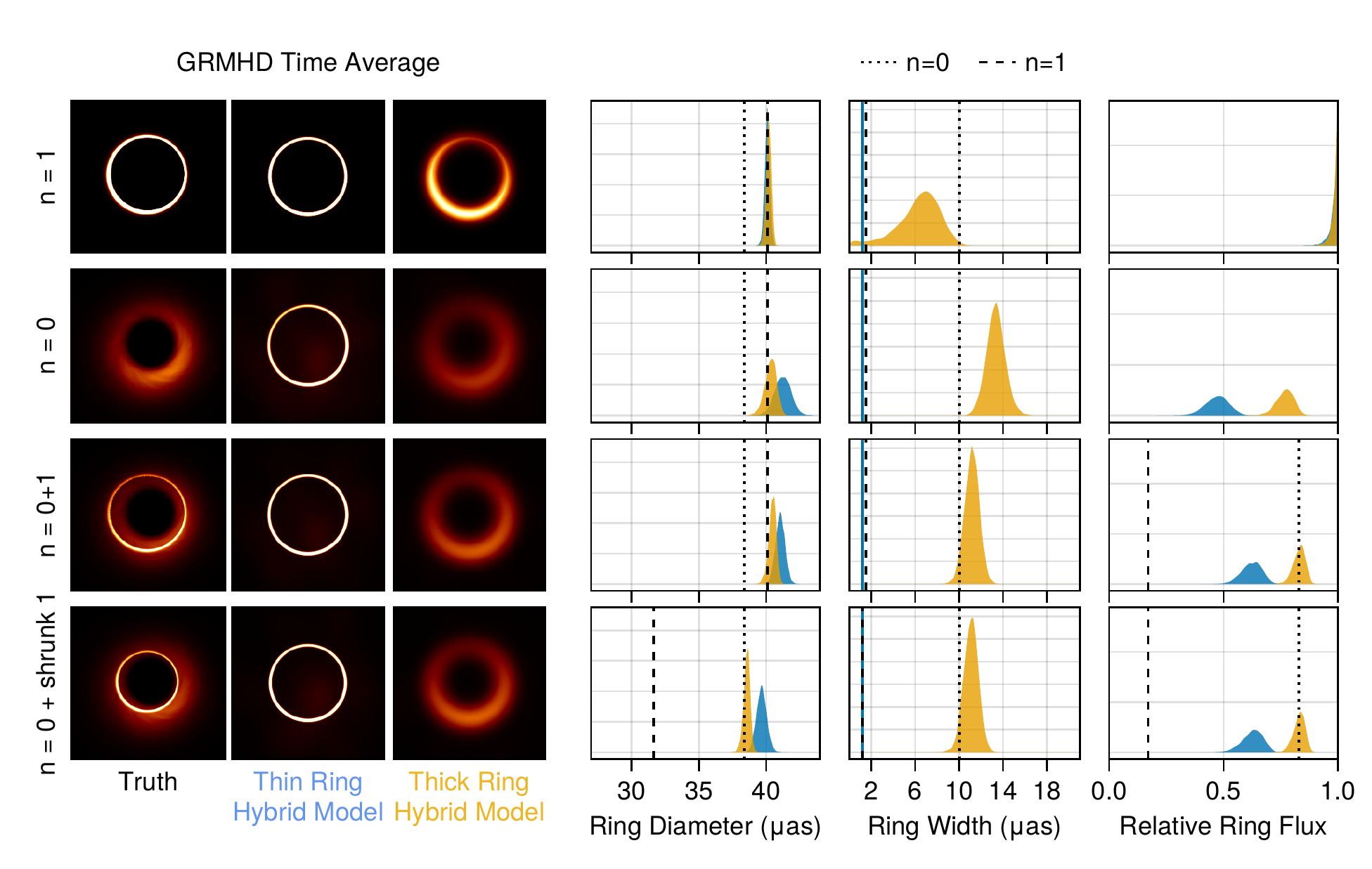}
    \caption{Results of fitting a time-averaged GRMHD simulation with coverage corresponding to the EHT~2017 array, following the same procedure and format of \autoref{fig:hybrid_fits}. The GRMHD simulation has a spin of $0.5$, inclination of $163^\circ$, $r_{\rm high} = 20$, and is in the MAD accretion state. %The temporal average was created by 
    }
    \label{fig:hybrid_fits_averaged}
\end{figure}

From \autoref{fig:hybrid_fits}, it appears that the thin ring component is modeling the combined emission of the $n=0$ and $n=1$ photon rings. One reason for this could be the fact that the $n=0$ emission has substantial small-scale ($< 5\,\muas$) structure due to plasma turbulence. This structure could be causing the thin ring to fit both the $n=0$ and $n=1$ structures. To assess whether this is the case, we then repeated the above analysis, but we replaced the GRMHD snapshot with a time-averaged GRMHD simulation. By time-averaging, we have averaged over the small-scale turbulence and created a smooth image $n=0$, yielding a more natural scale separation between the $n=0$ and $n=1$ emission that may make hybrid imaging more successful. Nevertheless, we find similar results when fitting the time-averaged images as those for the snapshots (see \autoref{fig:hybrid_fits_averaged}). Namely, the $n=0$ and $n=0+1$ fits give similar marginal posteriors for the thin ring diameter, and they are not substantially changed from the thick ring diameters. Additionally, the $n=0+ {\rm shrunk} 1$ fits also show that the measured ring diameter is substantially biased towards the $n=0$ emission diameter.

\subsection{Hybrid Imaging with the ngEHT}
\label{sec:hybrid_model_ngeht}

\begin{figure}[t]
    \centering
    \includegraphics[width=\textwidth]{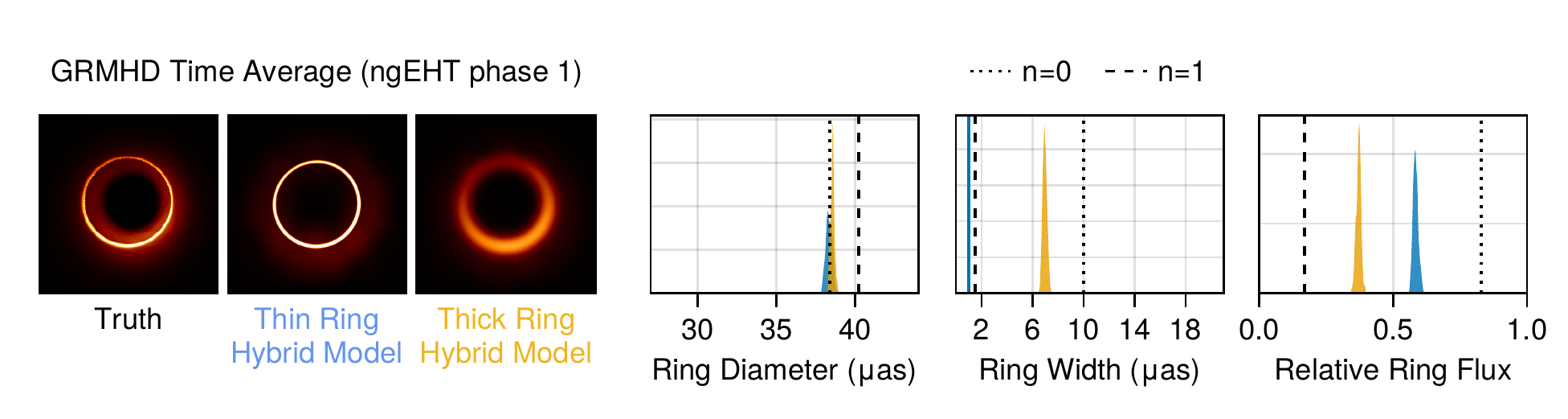}
    \caption{
    Results of fitting a time-averaged GRMHD simulation with no large-scale jet using a thick ring hybrid model and baseline coverage corresponding to the ngEHT~phase~1 array.}
    \label{fig:hybrid_densities}
\end{figure}

We now explore the prospects for hybrid imaging of \m87 with the ngEHT, focusing on the ngEHT~phase~1 array. We use a different GRMHD simulation for these tests, selecting one with no large-scale jet (see \autoref{fig:hybrid_densities}). Because the ngEHT has many short baselines, using an image with a prominent jet would significantly increase the necessary field of view and, hence, the number of raster elements required. Additionally, we only use a time-averaged simulation, since we expect hybrid imaging to perform best after time-averaging (see \autoref{ssec:EHT2017}). Finally, as for the geometric models, we assume a flat spectral index between 230~GHz and 345~GHz so that these bands can be easily combined in the modeling. 

The simulated dataset was created using the \texttt{ngehtsim} package using the same settings described in \autoref{ssec:geom_setup}. The model we fit matches \autoref{ssec:EHT2017}, except that we use a $13 \times 13$ raster with a $110~\muas$ field of view, which corresponds to an $8~\muas$ pixel size. We again fit the data using only closure data products, flagging any that have ${\rm SNR} < 3$, and we use a similar sampling strategy as for the geometric models in \autoref{sec:geom_modeling}. Because of the computational expense of this test, we only fit data for the full GRMHD image rather than examining the four decompositions into specific subimages (\autoref{fig:hybrid_fits_averaged}). As in \autoref{ssec:EHT2017}, we explore the prospects for both detection and measurement of the photon ring by fitting both the thick ring hybrid model and the thin ring hybrid model. Note that we include the ring ellipticity $\tau$ as a model parameter when fitting ngEHT synthetic data due to the improved baseline coverage.

\begin{figure}[!t]
    \centering
    \includegraphics[width=\textwidth]{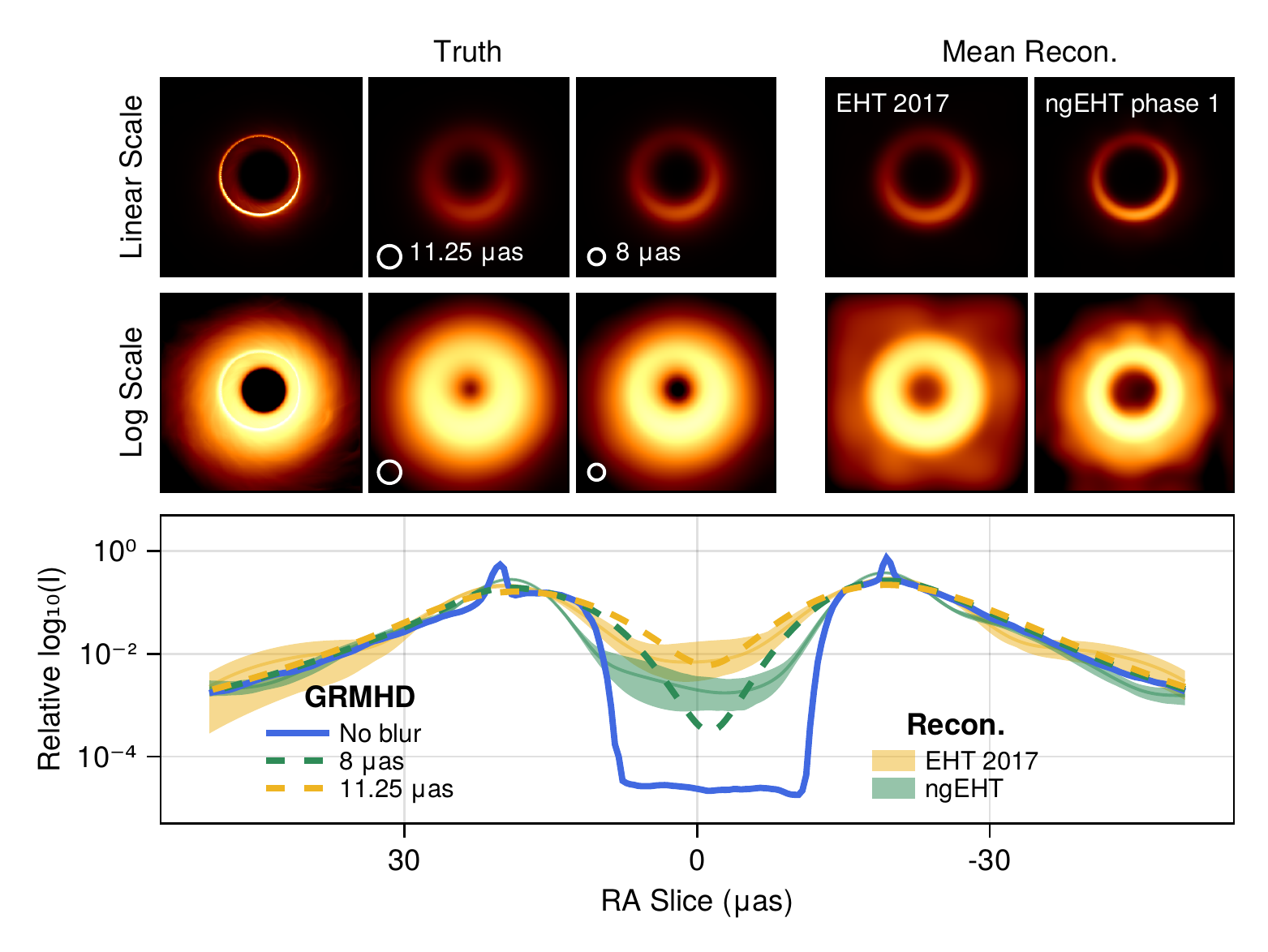}
    \caption{Summary of applying the hybrid model to fit both EHT~2017 and ngEHT~phase~1 synthetic data for the time-averaged GRMHD simulation shown in \autoref{fig:hybrid_fits_averaged}. The top row shows the ground truth image with different degrees of blurring (left group) and mean reconstructions using the thick ring hybrid model (right group) on a linear scale. The second row is the same set of images but plotted on a logarithmic scale. The $11\,\muas$ and $8\,\muas$ blurring kernels applied to the GRMHD simulation were chosen to match the raster resolutions of the hybrid models used for the EHT and ngEHT reconstructions, respectively. The bottom row shows the emission profile along the x-axis for the various images in the top two rows. Bands for the reconstructions denote $95\%$ credible intervals (including both the raster and ring components).} 
    \label{fig:grmhd_sim}
\end{figure}

\begin{figure}[!t]
    \centering
    \includegraphics[width=\textwidth]{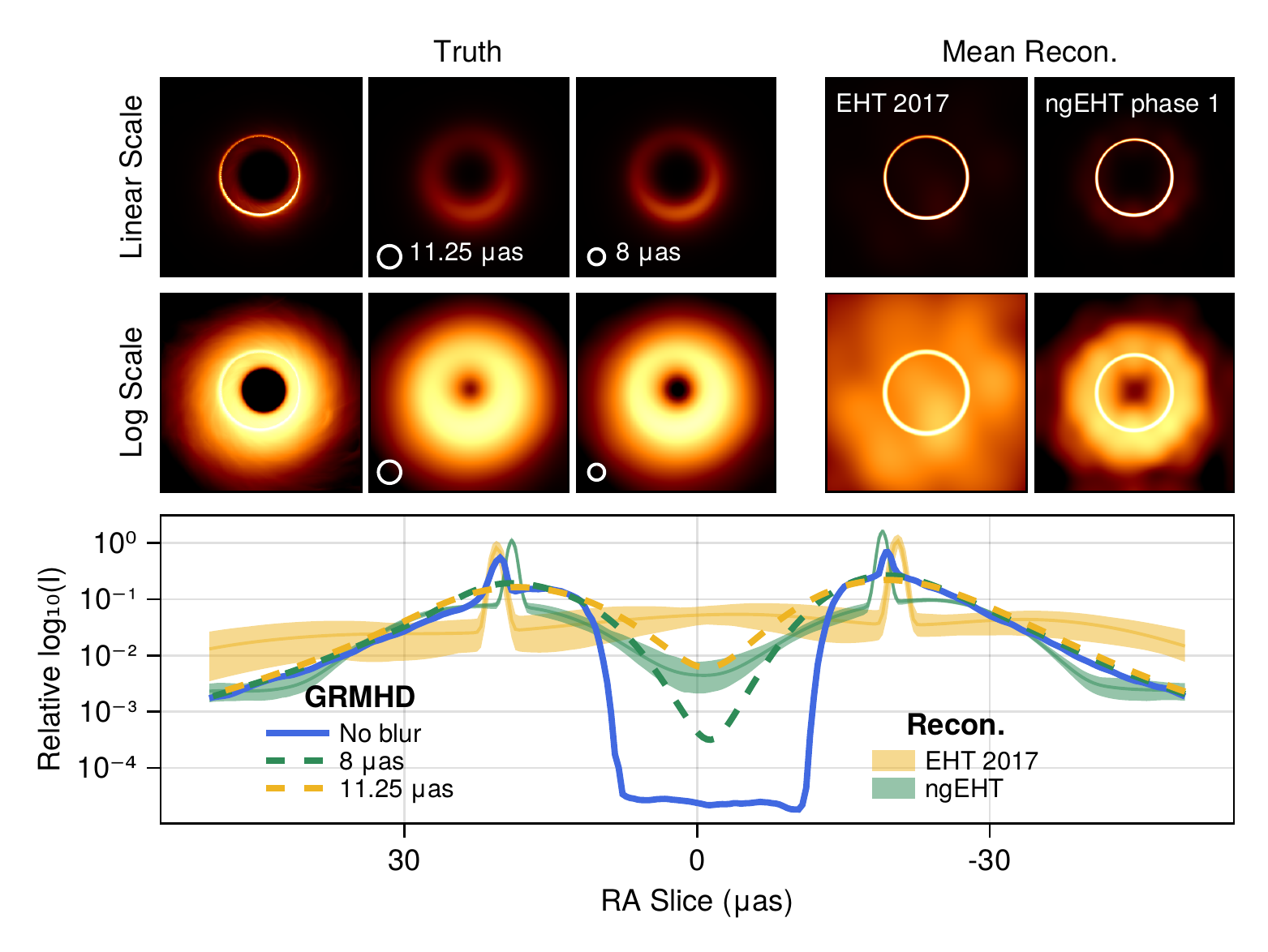}
    \caption{Similar to \autoref{fig:grmhd_sim} except fitting the thin ring hybrid model. In this case, the profiles show localized peaks associated with the thin ring, but they are displaced from the profile peaks of the $n=1$ photon ring. 
    } 
    \label{fig:grmhd_sim_thin_ring}
\end{figure}

The results are shown in \autoref{fig:hybrid_densities}. Unfortunately, the additional coverage of the ngEHT does not improve the biases seen in \autoref{ssec:EHT2017}. Nearly all ring parameters indicate that they are a combination of the $n=0$ and $n=1$ photon ring properties; the single exception is the measured ring diameter, which is consistent with the $n=0$ photon ring diameter for both the thin and thick ring fits. Unlike the EHT fits, the ring width is tightly constrained; however, it is much larger than the $n=1$ ring width and smaller than the $n=0$ ring width. The recovered fractional flux in the ring component is $0.36\pm 0.01$, while the true value for the $n=0$ and $n=1$ photon rings are $0.83$ and $0.17$, respectively. 

Comparing the ngEHT hybrid imaging results to the $n=0+1$ 2017 EHT results in \autoref{fig:geom_recon} we find that the results are somewhat similar, with the major difference being the posterior concentration for both the raster and ring parameters. In particular, rather than identifying a new solution mode, the posterior of the ring diameter and width for the ngEHT~phase~1 array results appear to be subsets of the EHT~2017 array results. Thus, even with the significant improvements of the ngEHT resolution, baseline coverage, and sensitivity, the hybrid imaging methodology does not successfully isolate the $n=1$ photon ring. 

\autoref{fig:grmhd_sim} and \autoref{fig:grmhd_sim_thin_ring} show the mean reconstructions and horizontal cross sections for the thick ring hybrid model and for the thin ring hybrid model, respectively. The cross sections demonstrate the significant improvement in image fidelity and dynamic range for the ngEHT relative to the EHT; all reconstructions robustly identify a central brightness depression, which is predominantly caused by the deep ``inner shadow'' in the simulated image \citep{Chael_2021}. However, no reconstructions for either model indicate a ring component that corresponds directly to the $n=1$ photon ring.

\section{Discussion}\label{sec:discussion}

We have explored the prospects for detecting and measuring the properties of the photon ring in \m87 using VLBI. Specifically, we have performed two types of tests using synthetic VLBI data, both within a Bayesian modeling framework implemented in the open-source library \comrade. The first type is simple geometric ``self-fits,'' which are idealized but easily interpreted. These geometric fits can be used to define firm requirements to detect the photon ring and to quantify how different array design choices affect the accuracy of photon ring parameter measurements. The second type generates synthetic data from more realistic GRMHD models and fits them in a ``hybrid imaging'' framework that simultaneously models a raster grid (similar to conventional VLBI imaging) and a geometric ring component. 

For the geometric tests, we find that the EHT baseline coverage and sensitivity cannot distinguish the direct ($n=0$) and secondary ($n=1$) emission (see also \citep{Lockhart_2022}). In particular, the width of the $n=1$ component is weakly constrained. However, the addition of longer baselines and higher observing frequencies in the simulated ngEHT coverage allows a firm detection of the $n=1$ photon ring. Thus, this test successfully provides minimal requirements for a photon ring detection with the ngEHT. 

For the more realistic GRMHD tests, we have demonstrated that the Bayesian VLBI modeling package \comrade can readily support posterior estimation using the hybrid imaging methodology with large rasters, even with ngEHT baseline coverage. Note that, our results are more nuanced and are strongly tied to limitations of the hybrid imaging methodology and of the VLBI data fitted. For our tests with EHT and ngEHT coverage, we find that
\begin{itemize}
    \item {\bf Hybrid imaging is prone to false positive detections of the photon ring.} Tests using images that only have direct ($n=0$) emission still show a strong preference for a ring component, even if the ring is restricted to be narrow. 
    \item {\bf Assuming a thin ring does not appreciably affect the other inferred ring parameters.} While the physically motivated assumption that the $n=1$ ring is narrow could plausibly affect the success of hybrid imaging, our fits are only weakly affected by this assumption.
    \item {\bf The fitted ring parameters in hybrid imaging do not correspond to the $n=1$ photon ring in the presence of confounding $n=0$ emission.} In our tests, the ring flux, width, and diameter are all affected by both the $n=0$ and $n=1$ emission and are generally most consistent with properties of the direct emission. 
\end{itemize}
In short, our tests indicate that estimates of black hole properties that rely on a rigid association of the ring component in hybrid imaging with the $n=1$ photon ring, such as those presented in \citetalias{broderick_photon_2022}, should be regarded with caution. A statistically significant detection of a thin ring component in a hybrid imaging model fit does not by itself demonstrate the existence of a photon ring in the source, because of the false positive tendencies of the method. Moreover, at minimum, mass-to-distance posteriors derived using hybrid imaging require an additional systematic error budget to account for the unknown bias from $n=0$ emission.

Our results highlight the challenge of estimating photon ring parameters in the superresolution regime, even when the modeling is informed by knowledge of the true image. Future studies, including blind testing within frameworks such as the ngEHT analysis challenge \citep{Roelofs_2022},\footnote{\url{https://challenge.ngeht.org/}} will provide additional guidance on what inferences are reliable. To convincingly make photon ring detections and measurements with real data, it will be imperative to demonstrate frequency and temporal independence of the inferred black hole parameters. Black hole images have strong dependence on frequency because of changing optical depth and synchrotron emissivity, so we expect that independent but consistent inferences across the full ngEHT frequency range (86-345\,GHz) will provide the most compelling empirical tests.  

We have explored a narrow range of possible implementations of hybrid imaging, examining only the difference between assuming a thick or thin ring component. Additional studies should explore the role of ring ellipticity, diameter, and relative flux density, all of which could use physically informed priors or information from complementary observations (e.g., from resolved stellar orbits of \sgra).   

Future studies that explore other observational signatures of the photon ring, such as those in linear polarization \citep[e.g.,][]{Himwich_2020,Palumbo_2022_pol}, in circular polarization \citep[e.g.,][]{Moscibrodzka_2021,Ricarte_2021}, and in the time domain \citep[e.g.,][]{Moriyama_2019,Tiede_2020,Hadar_2021,Wong_2021}, can provide important pathways to detection and measurement, as well as creating additional validation opportunities across data products and analysis methods. Finally, while we have focused our tests on \m87, \sgra is another target for millimeter VLBI for which photon ring detection may soon be possible. \sgra has a somewhat larger angular gravitational radius than \m87, so its photon ring is likely to be larger as well. Unlike \m87, \sgra has an exquisitely measured mass from resolved stellar orbits \citep[e.g.,][]{Do_2019,Gravity_2019} which can either be integrated as an informative prior or can be used as a powerful consistency test on photon ring inferences. However, \sgra has the additional challenges of strong interstellar scattering \citep[e.g.,][]{Bower_2006,Psaltis_2018,Johnson_2018,Issaoun_2019} and rapid variability \citep[e.g.,][]{Wielgus_2022,SgrAEHTCIV}, and we expect that the requirements to detect the photon ring in \sgra may be more stringent than those for \m87. 

%%%%%%%%%%%%%%%%%%%%%%%%%%%%%%%%%%%%%%%%%%
\vspace{6pt} 

%%%%%%%%%%%%%%%%%%%%%%%%%%%%%%%%%%%%%%%%%%
\authorcontributions{Conceptualization, P.T., M.D.J, and D.W.P.; methodology, P.T. and M.D.J.; software, P.T.; writing---original draft preparation, P.T., M.D.J., and D.W.P.; writing---review and editing, P.T., M.D.J., D.W.P., D.C.M.P., D.O.C., and P.G.; All authors have read and agreed to the published version of the manuscript.}

\funding{This work was supported by the Black Hole Initiative at Harvard University, which is funded by grants from the John Templeton Foundation and the Gordon and Betty Moore Foundation to Harvard University. This work was also supported by National Science Foundation grants AST-1935980 and AST-2034306 and the Gordon and Betty Moore Foundation (GBMF-10423).}

\acknowledgments{We thank Avery Broderick, Boris Georgiev, and Britt Jeter for many helpful conversations about hybrid imaging.}

\conflictsofinterest{``The authors declare no conflict of interest.''} 

%%%%%%%%%%%%%%%%%%%%%%%%%%%%%%%%%%%%%%%%%%
\begin{adjustwidth}{-\extralength}{0cm}

\reftitle{References}

\bibliography{references}

\end{adjustwidth}
\end{document}